# Zipf's Law for All the Natural Cities in the United States: A Geospatial Perspective

Bin Jiang[(1)] and Tao Jia[(2)]

[(1)] Department of Technology and Built Environment, Division of Geomatics
University of Gävle, 801 76 Gävle, Sweden
Email: bin.jiang@hig.se,

[(2)]Future Position X, Box 975, 801 33 Gävle, Sweden
Email: jiatao83@163.com

**Abstract**
This paper provides a new geospatial perspective on whether or not Zipf's law holds for all cities or for the largest cities in the United States using a massive dataset and its computing. A major problem around this issue is how to define cities or city boundaries. Most of the investigations of Zipf's law rely on the demarcations of cities imposed by census data, e.g., metropolitan areas and census-designated places. These demarcations or definitions (of cities) are criticized for being subjective or even arbitrary. Alternative solutions to defining cities are suggested, but they still rely on census data for their definitions. In this paper we demarcate urban agglomerations by clustering street nodes (including intersections and ends), forming what we call natural cities. Based on the demarcation, we found that Zipf's law holds remarkably well for all the natural cities (over 2-4 million in total) across the United States. There is little sensitivity for the holding with respect to the clustering resolution used for demarcating the natural cities. This is a big contrast to urban areas, as defined in the census data, which do not hold stable for Zipf's law.

**Keywords:** Natural cities, power law, data-intensive geospatial computing, scaling of geographic space

## 1. Introduction
The size distribution of cites in a country or across the world demonstrates a striking regularity known as Zipf's law, so named after George Zipf (1949) who first popularized the empirical finding. It was Auerbach (1913) who first discovered that the city size distribution could be approximated by a power law distribution. If we tabulate all the cities of a country and rank them according to their size, e.g. by population, the first largest city is twice as big as the second largest, three times as big as the third largest, and so on. To put it another way, the size of a city is inversely proportional to its rank. This indicates a dual aspect of Zipf's law: the size distribution follows a power law, and the exponent or the Zipf value is close to 1.0. This is different from how we usually observe measurements, i.e. values are frequently centered around an average value. For instance, there is a mean height for men and women. In the case of city sizes, the average population of cities is not a useful measurement. Rather, there are far more small cities (smaller than the average) than large ones (larger than the average); usually about 90% of cities are characterized as small cities, while about 10% of cities as big cities.

Underlying Zipf's law, there are two fundamental issues that have occupied researchers from physics, economics, linguistics, and geography for several decades. The first is whether or not Zipf's law holds for different countries or regions. The commonly held opinion is yes. For example, the size distribution of larger cities in the United States fairly well fits the power law with an exponent close to 1.0; amazingly this regularity has been held for nearly a century. However, some researchers argue that Zipf's law holds only in the upper tail, or for the largest cities, and that the size distribution of cities follows alternative distributions (e.g. lognormal distributions, stretched exponential distributions) other than a power law (e.g., Gabaix and Ioannides 2004, Batty 2006, Benguigui and Blumenfeld-Lieberthal 2007, Laherrere and Sornette 1998). The second issue is to provide an explanation as to why and how this simple regularity has emerged (Cordoba 2008). The latter issue is particularly baffling and intriguing, as noted by Paul Krugman (1996) "*we have complex messy models,* *yet reality is startlingly neat and simple"*. Existing urban economics models fail to offer a sound explanation. For example, the random grown model proposed by Simon (1955) can explain a power law, but it is hard to reproduce one with the right exponent of 1.0. This paper contributes mainly to the first issue, although it may add hints or possibilities to tackling the second issue.

A major problem with the first issue is how to define a city or demarcate a city boundary. A traditional way is to take cities as defined in the US census data, e.g., metropolitan areas or census-designated places (Eeckhout 2004, Chen 2010). Cities defined in the traditional way are somewhat arbitrary due to their legally or administratively determined boundaries. Not all people live in these "cities". This is because some places are excluded from being census places due to state law. According to the US 2000 census data, only 80% of the entire USA population lived in the 276 metropolitan areas, while only 74% live in the 25,359 census places. Recently, new approaches have emerged to objectively define cities or city boundaries. For example, Holmes and Lee (2009) defined cities as individual cells bounded by six-by-six-mile grids. The city size is then measured by the population of the populated sites within the grid city boundaries. Rozenfeld et al. (2009) adopted a city clustering algorithm to automatically derive city boundaries by clustering populated sites with a prescribed distance, e.g., 3000 meters. These recent efforts abandoned the city definitions imposed by the census data, but they still rely on populated sites (from the census data) for demarcating or defining city boundaries. The populated sites are individual locations representing census tract codes. The city boundaries derived by the new approaches might sound more reasonable, but they still suffer from the same problem, i.e., not all population concentrations are counted.

We propose an approach that includes all cities or human settlements for evaluating Zipf's law. We define cities based on street nodes (including intersections and ends) without incorporating any census information. This is a natural way – a bottom up approach to defining cities, so the cities are called natural cites. The definition of the natural city is based on the fact that human activities are constrained to streets – no streets no human activity, or alternatively no street nodes no residential places or cities. This way we can extract all places from the largest megacity with 8 million residents to the smallest town with a single person. This is practically possible, since the smallest town must have at least one street junction or street end. Without incorporating any census information, our approach shows little bias imposed by the census data.

The remainder of this paper is organized as follows. Section 2 describes two sets of data: (a) street nodes and clustered natural cities from the street nodes using massive OpenStreetMap (OSM) data, and (b) urban areas and population from the census data. Section 3 introduces power law distributions and how to detect a power law when there is one. In Section 4, we report our findings on the examination of Zipf's law for the natural cities. Section 5 discusses the contributions of this paper. Finally, Section 6 concludes the paper and points to future work.

**2. Data**
Two datasets of the contiguous USA, Alaska and Hawaii are involved in the study. The first dataset consists of street nodes (about 25 million) and the derived natural cities (2-4 million) by clustering the street nodes. The second data set is taken from the US census including 3,638 urban areas and population assigned to 65,997 population centers. The use of the massive street nodes and the clustered natural cities is a novel aspect of this study. We used the second dataset mainly for a comparison purpose.

**2.1 Street nodes and natural cities**
The main data used in the study are the street nodes of the entire country. This is the primary data, on which we derived the natural cities using the city clustering algorithm (Rozenfeld et al. 2009). It is important to note that human activities are constrained to streets: no street no human activity. To put it more precisely, there would be no human activity if there were no street nodes. The pattern of street nodes reflects to a great extent that of human settlements. The smallest human settlement (e.g., with one house) needs at least one street node. Street nodes include both street junctions and ends. Junctions are the intersections of two or multiple streets on the same plane. The process of extracting street junctions is fairly simple and straightforward. Any street node with more than three street segments is considered to be a street junction. It should be noted that highway bridges are excluded from being a junction, as they are not crossed on the same plane. This applies to situations where two highways at two different planes are intersected by links. This is clearly tagged in the OSM database. We wrote a little program to extract 24,657,017 street nodes for the entire country from over 120 gigabytes of street network data.

Unlike legally defined cities such as metropolitan areas or census places that are imposed from the top down, the natural cities are defined from the bottom up. We apply the city clustering algorithm to the massive street nodes to obtain individual urban agglomerations or natural cities. This process is described in the following recursive function:

```
Select any street node as current point;
Recursive Function Agglom (current point)
    Draw a circle with radius around the current point;
    Search other points within the circle, and add to a point set;
    If (the point set = empty) Then
        Return;
    Else
        Pick up any point from the point set as the current point;
        Remove the current point from the point set;
    Call Function Agglom (current point)
```

The radius in the above function is what we call the clustering resolution. The final result of natural cities relies on the clustering resolution, i.e., the finer the resolution, the more natural cities. In fact, this clustering algorithm is based on a simple measurement of location similarity or nearest neighbor analysis (Jacquez 2008). If we set the resolution as 1 meter for example, the number of natural cities would be the same as that of street nodes. This is because there are never two street nodes within 1 meter. Usually the resolutions should be about the size of street blocks, e.g., > 300 meters. Therefore, we chose four resolutions: 400 m, 500 m, 600 m and 700 m for our investigation. The size of natural cities can be measured by the number of clustered street nodes. The size can also measured by the areas of individual natural cities. For this purpose, we need to delimit city boundaries. The delimitation of city boundaries is based on a raster approach by imposing a grid on top of the urban agglomerations. Those cells containing street nodes are set to 1 and others to 0 – thus a binary map is created. Starting from an initial cell, we then traverse individual cells with value 1 in all directions until coming back to the starting cell; a boundary is thus formed. This process goes continuously until all boundaries are formed. Figure 1 shows a part of the natural cities from the region near New York.

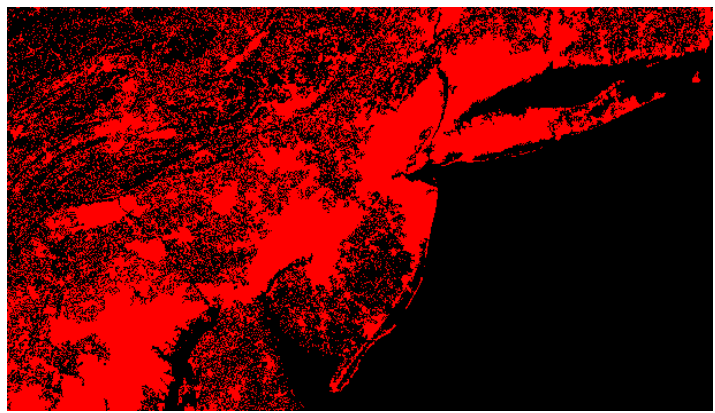

Figure 1: Natural cities near the New York region

### 2.2 Urban areas and population
Urban areas are one of the formally defined geographic areas in the census 2000 data. It consists of urbanized areas (with population > 50,000) and other urban entities (with population between 2,500 and 49,999). We downloaded the data from http://www.census.gov/geo/www/cob/ua2000.html. There are initially 11,880 urban areas, but many of them have the same names, or have no name at all. For those without a name, we merged them into large adjacent urban areas. For those with the same name, we also merged them into one unit. Eventually, after the merging processes, there are 3,638 urban areas represented as a polygon layer (*c.f.*, Figure 2).

Population data contain population information at the level of census tracts for individual population centers. There are a total of 65,997 population centers, each ranging from 1 to 36,146 people. The data were downloaded from http://www.census.gov/geo/www/cenpop/cntpop2k.html (excluding 307 invalid records because *x, y* and *pop* are all set to zero). We thought at the very beginning that this is the same dataset studied in Rozenfeld et al. (2009), but realized that they are different. This is because with their dataset the population ranges from 1500 to 8000 people for each record as described in the paper (Rozenfeld et al. 2009). However, the two datasets have the same format. Each entry of the data is uniquely identified by 11 digits, e.g., for the first entry "01001020100,1921,+32.47507,-86.486814". The first 2 digits correspond to the state, the next 3 to the county within the state and the rest to the census tract. The first record indicates some state (01), some county (001), and some census tract (020100), with ´population 1921 located at +32.47507, -86.486814. Overlapping urban areas and the population data, we found that there are many population centers that are not within any urban area. Statistical analysis indicates that only 49,114 population centers are within urban areas, and the centers account for 76% of the entire population.

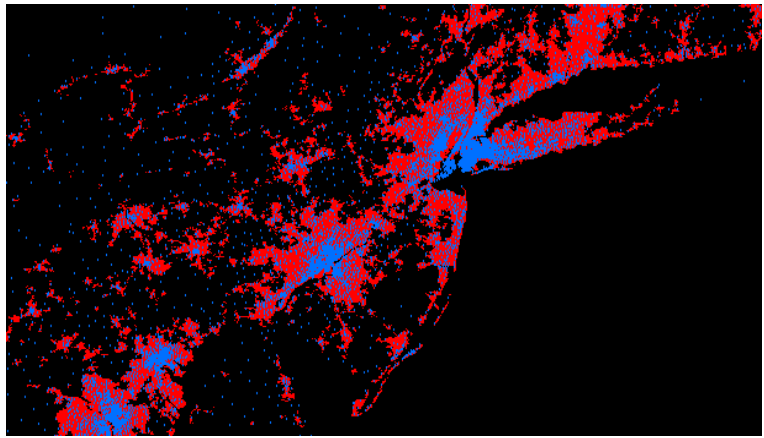

Figure 2: Urban Areas (red patches) and Population (blue points) near New York for illustration purposes

Urban areas as a proxy for city size and in contrast to city population have been used in some previous studies (e.g., Chen and Zhou 2008, Benguigui et al. 2006) for examining city structure and dynamics. We followed the same idea but used them as a reference data. The examined result serves as an important reference to this study, as well as to previous studies using metropolitan areas and census places.

### 3. The mathematics of cities - Zipf's law, power laws, and Pareto distributions

This section provides an introduction to power laws with a particular focus on two issues: how to detect a power law when there is one and how to compute the power law exponent. There is a vast literature on the subject over the past decades in a variety of disciplines, but confusion and misunderstanding surrounding the two issues have existed for a long time until very recently (e.g., Adamic 2002, Newman 2005). Thanks to the availability of massive data collected about and through the Internet, World Wide Web, and all kinds of social networks, this subject of power laws has gained an increasing importance and received a revival of interest. Our introduction is intended to be brief; a more detailed account can be found in the literature (e.g., Clauset et al. 2009).

Zipf's law when applied to the size distribution of cities is a typical nonlinear relation between city rank *(r)* and city size *(s)*, sometimes called rank-size rule. Formally it is expressed by,

$$s = r^{-1} \qquad [2]$$

It is clear that the city size *s* is 1, 1/2, 1/3… with respect to city rank 1, 2, 3…. This is what we mentioned at the beginning of this paper – the first largest city is twice as big as the second largest, and three times as big as the third largest, and so on. Generally, a power law is expressed by

$$y = kx^{-\alpha} \qquad [3]$$

where *x* is some quantity, both *k* is a constant, and $\alpha$ is the power law exponent.

This kind of power law is also known as a Pareto distribution after the Italian economist Vilfredo Pareto (1848 – 1923). Pareto was initially interested in the distribution of wealth in a country. He found that this distribution is very unequal, i.e., 20% of the people own 80% of the wealth, while 80% of the people own only 20% of wealth – thus the rich gets richer. In fact, the Pareto distribution or the 80/20 rule has been found in many other natural and man-made phenomena.

To detect a power law, we can do a plot at logarithmic scales to see if a straight line appears, i.e.,

$$\ln(y) = -\alpha \ln(x) + \ln(k) \qquad [4]$$

The method suffers from errors in the logarithmic tail of the distribution. The end of the logarithmic tail looks messy because each bin only has very few samples in it. One possible solution is to use varying widths of bins in the histogram, with each bin *b* being increased by $2^b$, to achieve a more homogeneous number of samples per bin. This can help reduce errors in the tail. This method can be further refined, e.g., by using the frequency per logarithmic bin normalized by dividing by bin width to get the probability density (e.g., Newman 2005, Viswanathan et al. 1999). This solution is actually based on the probability density function (PDF). Another solution, probably a much better one, is to use the cumulative density function (CDF), i.e., a plot of the probability that the quantity *x* is greater than or equal to a certain value. This form of the power law distribution is Zipf's law or the Pareto distribution.

Next we have to determine the power law exponent $\alpha$. This is usually done by using the least-squares fit, but it is known to introduce systematic bias (Goldstein et al. 2004). Because a power law distribution is likely to be confused with other heavy tail distributions, such as the lognormal distribution and the stretched exponential distribution, it is very tricky to make a power law hypothesis. Many more reliable methods have been suggested (Goldstein et al. 2004, Newman 2005), and they are based on the maximum likelihood methods and the Kolmogorov-Smirnov (KS) test for respectively identifying and quantifying power-law distributions. In other words, these methods can be used not only to fit a power law to data (or part of the data), but also for assessing how good the fit is in comparison with other heavy tailed distributions. The estimated exponent is given by,

$$\alpha = 1 + n \left[ \sum_{i=1}^{n} \ln \frac{x_i}{x_{\min}} \right]^{-1} \qquad [5]$$

where $\alpha$ denotes the estimated exponent, and $x_{\min}$ is the smallest value for which the power law holds. It should be noted that the exponent of Zipf's law is $\alpha - 1$.

A modified KS test suggested by Clauset et al. (2009) is adopted in this study to assess the goodness of fit, i.e., how good city sizes fit a power law distribution. A fundamental idea is the maximum distance ($\delta$) between the CDFs of the data and the fitted model:

$$\delta = \max_{x \geq x_{\min}} |f(x) - g(x)| \quad [6]$$

Where $f(x)$ is the CDF of the synthetic data with a value of at least $x_{\min}$, and $g(x)$ is the CDF for the power law model that best fits the data while $x \geq x_{\min}$.

With the fitted model $g(x)$, we generate 1000 synthetic datasets that follow a perfect power law above $x_{\min}$ but have the same non-power-law distribution as the observed data below, and recalculate the maximum distance between $f(x)$ and the fitted model, i.e., $\delta_i (i = 1, 2, \ldots 1000)$. A goodness of fit index p-value is defined by

$$p = \frac{\text{the number of } \delta_i \text{ whose values are greater than } \delta}{1000} \quad [7]$$

The p-value indicates to what extent the data fit the model. The larger the p value, the more significant is the model, and *p* values greater than 0.05 are considered to be acceptable for a goodness of fit. It is important to note that the way of detecting a power law distribution and the related KS test are a recent advance. They are particularly useful in detecting a power law distribution from other fat tail distributions. On the other side, computing p values can be very time consuming in particular for a big sample.

## 4. Results and discussion
In this section, we will report in detail our investigation about the validity of Zipf's law using natural cities. The city sizes are measured from both street nodes and physical areas. The results are put in comparison with those based on the urban areas, one of the city demarcations imposed by the US census. The sizes of urban areas are measured by population and physical areas.

### 4.1 The long tail of the distribution of natural cities
The first finding from our study is that there is a long tail for the distribution of natural cities. That is, a vast majority of natural cities are small cities, staying in the tail; while a minority of natural cities are big cities, staying in the head (Figure 3). Given an average size of all cities (m), the corresponding rank R(m) would partition all cities into two categories: those 10% bigger than the average in the head and those 90% smaller than the average in the tail. This is also shown in Table 1 where we provided the actual number of cities and their percentages with respect to four different clustering resolutions: 400 m, 500 m, 600 m, and 700 m. This result is very intriguing, indicating that the majority is trivial, while the minority is vital. This kind of imbalance is a good indicator of a power law distribution, which will be further examined in the following. This imbalance is often characterized by the 80/20 rule. It should be noted that the underlying meaning of the 80/20 rule is not how precise the 80% or 20% is, but rather the imbalance between the head and the tail. For example, we show it here to be around 90% and 10%, rather than 80% and 20%.

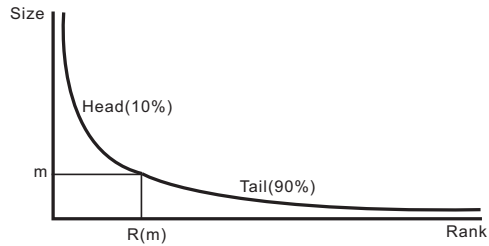

Figure 3: Illustration of a long tail distribution

Table 1: Number of natural cities with respect to four clustering resolutions

| Clustering resolutions | 700 | 600 | 500 | 400 |
|---|---|---|---|---|
| # of natural cities (all) | 2373382 | 2933849 | 3727129 | 4779305 |
| # of natural cities ( < mean) | 2208939 | 2706046 | 3391758 | 4342613 |
| Tail (< mean) | 93% | 92% | 91% | 91% |
| Head (> mean) | 7% | 8% | 9% | 9% |

**4.2 Zipf's Law for natural cities and urban areas**

The second finding is a further refined result from the first. We found that Zipf's law holds remarkably well for all the natural cities with respect to different resolutions: 400 m, 500 m, 600 m, and 700 m. Figure 4 shows a log-log plot, where the straight line stretches over more than two decades. Importantly, the power law exponent is around 2.0 (Table 2). This implies that the Zipf exponent is around 1.0. We deliberately choose different sets of natural cities from the biggest 150, 1500, 15000, 150000 and all the cities. Our investigation ends up with a very slight change of the power law exponent mostly at the second decimal. However, as we can see from Table 2 and Figure 5, the exponent based on urban areas is not stable with respect to different parts of the tail. They are not a truly scale free. More critically, the exponent is very different from 2.0; see columns namely Urban Areas in Table 2. It is important to note that the examination of Zipf's law is based on a rigorous statistical test – a modified KS test as introduced in the above section. The corresponding values are shown in Table 3. They are all greater than the threshold of 0.05, and some of them show an exceptional goodness of fit with a p-value equal to 1.0.

We tend to believe that 500 m and 600 m are the best resolution among the four options. This is based on both visual inspection in comparison with the related urban areas and a simple reasoning on the nature of the city clustering algorithm. A resolution 1000 m would merge most of real cities as one natural city. Resolution 700 is still too high to separate some cities near New York, although it seems like the best in terms of closeness to the value 2.0 in Table 2 and the highest p values shown in Table 3. Eventually we end up with the two suggested best resolutions.

Table 2: Power law exponent for natural cities in comparison with urban areas

| | 700 | | 600 | | 500 | | 400 | | Urban Areas | |
|---|---|---|---|---|---|---|---|---|---|---|
| Biggest cities | Nodes (α) | Areas (α) | Nodes (α) | Areas (α) | Nodes (α) | Areas (α) | Nodes (α) | Areas (α) | Pop. (α) | Areas (α) |
| 150 | 1.98 | 2.09 | 2.00 | 2.10 | 2.06 | 2.14 | 2.14 | 2.22 | 1.91 | 2.08 |
| 1500 | 2.01 | 2.09 | 2.03 | 2.10 | 2.06 | 2.10 | 2.10 | 2.12 | 1.74 | 1.81 |
| 15000 | 2.01 | 2.09 | 2.04 | 2.11 | 2.06 | 2.12 | 2.11 | 2.15 | NA | NA |
| 150000 | 2.01 | 2.09 | 2.04 | 2.11 | 2.06 | 2.12 | 2.11 | 2.15 | NA | NA |
| All cities | 2.01 | 2.09 | 2.04 | 2.11 | 2.06 | 2.12 | 2.11 | 2.15 | 1.74 | 1.8 |

Table 3: P values from KS test for natural cities in comparison with urban areas

| | 700 | | 600 | | 500 | | 400 | | Urban Areas | |
|---|---|---|---|---|---|---|---|---|---|---|
| Biggest cities | Nodes (p) | Areas (p) | Nodes (p) | Areas (p) | Nodes (p) | Areas (p) | Nodes (p) | Areas (p) | Pop. (p) | Areas (p) |
| 150 | 0.29 | 0.81 | 0.32 | 0.77 | 0.36 | 0.17 | 0.07 | 0.30 | 0.24 | 0.01 |
| 1500 | 0.95 | 0.17 | 0.67 | 0.18 | 0.99 | 0.55 | 0.39 | 0.50 | 0.01 | 0.00 |
| 15000 | 0.85 | 0.11 | 0.83 | 0.45 | 0.92 | 0.21 | 0.34 | 0.29 | NA | NA |
| 150000 | 0.84 | 0.11 | 0.83 | 0.44 | 0.94 | 0.22 | 0.40 | 0.27 | NA | NA |
| All cities | 0.84 | 0.10 | 0.90 | 0.41 | 0.88 | 0.18 | 0.46 | 0.28 | 0.10 | 0.03 |

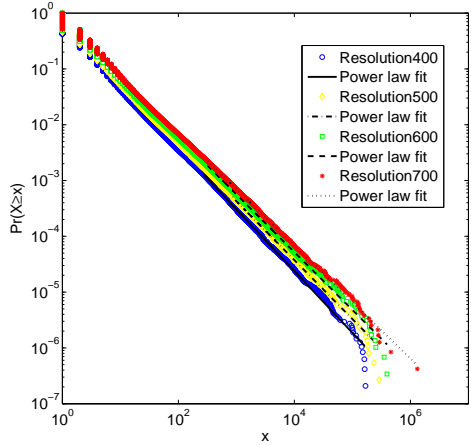

Figure 4: (Color online) Power law distribution with respect to different clustering resolutions
(Note: size x is measured by the number of nodes)

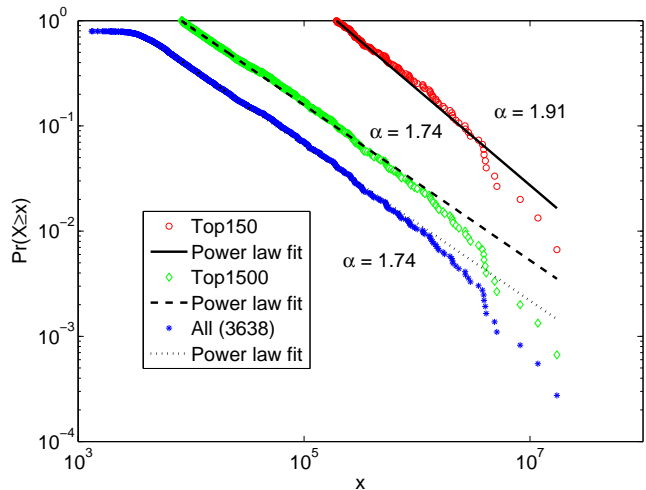

Figure 5: (Color online) Power law distribution of urban areas with respect to different parts
(Note: size x is measured by the urban areas)

**4.3 Zipf's law for individual states**
We go further down to individual states to examine Zipf's law. A power law holds remarkably well for all natural cities within individual states, but the exponent deviates from 2.0 (or Zipf exponent 1.0) and it varies from state to state. We choose a resolution size of 600 as an example, and the results are shown in Table 4. This result indicates that Zipf's law for natural cities is different from one state to another. To reiterate, Zipf's law has a dual aspect: the power law distribution on the one hand and Zipf exponent close 1.0 on the other (or equivalently power law exponent close to 2.0). This striking regularity is what Paul Krugman (1996) refer to the startlingly simple reality. In this sense, we can claim that Zipf's law is not universal for different geographical regions or countries in terms of the exponent. Intuitively, we understand from a mathematical point of view that the bigger the alpha values, the more heterogeneous the city sizes. What does that mean in terms of economic or other aspects? Does the fact that some states have the same exponent suggest common economic and other characteristics? Such questions are beyond the scope of this paper, but are surely of interest to be addressed by the science community.

**5. Contributions of this paper**
It is important to note that there are many other methods to derive city boundaries from street nodes. For example, Borruso (2003) studied a couple of methods, one of which is based on kernel density estimation from street junctions, and then demarcates city boundaries based on the density. This is simple enough. However, this particular method involves too many parameters such as resolution, bandwidth and density cutoff that make the city boundaries too sensitive. The method we adopted in this paper involves only resolution if the city sizes are measured by the number of street nodes. However, the methods adopted by this study are not novel aspect of this paper. In the same way, the method of detecting a power law and estimating the exponent is not new either, and it is well documented in the literature. The novel aspects of this paper lie in the data and the findings. First of all, we processed over 120 gigabytes of street network data in order to get the 25 million street nodes. Eventually we got over dozens of million of natural cities based on data-intensive computing. Although we rely on a state-of-the art personal computer, we exhausted all the computing capacity, taking days of computer time to get the work done. Secondly the findings and insights about Zipf's law constitute another novel aspect of this paper. The validity of Zipf's law has never been examined at such a massive data level and from such a unique geospatial perspective. We believe that our study in particular this unique geospatial perspective contributes to the long time debate about city size distributions from an array of disciplines.

Table 4: Power law exponent and p values for natural cities within individual states

| States | Nodes | | Area | | States | Nodes | | Area | |
|---|---|---|---|---|---|---|---|---|---|
| | α | p | α | p | | α | p | α | p |
| Alabama | 1.95 | 0.95 | 2.12 | 0.15 | Montana | 2.08 | 0.06 | 2.34 | 0.66 |
| Alaska | 2.28 | 0.12 | 2.48 | 0.90 | Nebraska | 2.16 | 0.77 | 2.34 | 0.33 |
| Arizona | 1.98 | 0.00 | 2.28 | 0.02 | Nevada | 1.88 | 0.87 | 2.11 | 0.55 |
| Arkansas | 1.99 | 0.79 | 2.12 | 0.47 | New Hampshire | 1.96 | 0.85 | 2.14 | 0.19 |
| California | 1.83 | 0.57 | 1.98 | 0.19 | New Jersey | 1.80 | 0.29 | 1.77 | 0.55 |
| Colorado | 1.93 | 0.02 | 2.30 | 0.43 | New Mexico | 2.01 | 0.64 | 2.14 | 0.11 |
| Connecticut | 1.89 | 0.32 | 1.99 | 0.07 | New York | 2.04 | 0.21 | 2.17 | 0.00 |
| Delaware | 1.85 | 0.01 | 1.86 | 0.75 | North Carolina | 1.95 | 0.00 | 1.82 | 0.28 |
| Florida | 1.82 | 0.00 | 1.90 | 0.82 | North Dakota | 2.34 | 0.65 | 2.52 | 0.11 |
| Georgia | 1.91 | 0.37 | 2.02 | 0.96 | Ohio | 1.96 | 0.29 | 2.00 | 0.82 |
| Hawaii | 1.60 | 0.03 | 1.83 | 0.12 | Oklahoma | 2.05 | 0.26 | 2.20 | 0.33 |
| Idaho | 2.11 | 0.20 | 2.18 | 0.12 | Oregon | 1.93 | 0.33 | 2.10 | 0.03 |
| Illinois | 2.05 | 0.43 | 2.16 | 0.46 | Pennsylvania | 1.93 | 0.29 | 2.02 | 0.95 |
| Indiana | 1.99 | 0.82 | 2.09 | 0.87 | Rhode Island | 1.89 | 0.66 | 2.09 | 0.18 |
| Iowa | 2.12 | 0.39 | 2.16 | 0.60 | South Carolina | 1.91 | 0.84 | 2.16 | 0.03 |
| Kansas | 1.99 | 0.16 | 2.30 | 0.66 | South Dakota | 2.21 | 1.00 | 2.57 | 0.03 |
| Kentucky | 1.94 | 0.43 | 2.08 | 0.49 | Tennessee | 1.91 | 0.78 | 2.04 | 0.76 |
| Louisiana | 1.91 | 0.43 | 2.15 | 0.06 | Texas | 2.07 | 0.43 | 2.24 | 0.46 |
| Maine | 2.10 | 0.72 | 2.31 | 0.15 | Utah | 2.06 | 0.39 | 2.33 | 0.55 |
| Maryland | 1.84 | 0.06 | 1.79 | 0.81 | Vermont | 2.17 | 0.54 | 2.38 | 0.18 |
| Massachusetts | 1.87 | 0.03 | 1.85 | 0.99 | Virginia | 2.01 | 0.26 | 2.06 | 0.02 |
| Michigan | 2.15 | 0.62 | 2.29 | 0.43 | Washington | 1.98 | 0.11 | 2.15 | 0.83 |
| Minnesota | 2.29 | 0.82 | 2.49 | 0.95 | West Virginia | 1.94 | 0.81 | 1.95 | 0.76 |
| Mississippi | 2.01 | 0.83 | 2.24 | 0.08 | Wisconsin | 1.99 | 0.02 | 2.23 | 0.25 |
| Missouri | 1.99 | 0.16 | 2.16 | 0.26 | Wyoming | 2.07 | 0.21 | 2.41 | 0.35 |

Having stated the novel aspects of this paper, it is worthwhile to briefly mention the OSM data from which street nodes were extracted. OSM is a wiki-like collaboration, or grassroots movement, to create a free editable map of the world using data from portable GPS units, aerial photography, and other free sources (*c.f.*, Haklay and Weber 2008). It has become a phenomenal success as one of the best examples of volunteered geographic information contributed by individuals and supported by Web 2.0 technologies (Goodchild 2007). Both the OSM data size and the geographic coverage have been growing dramatically over the past few years, in large part thanks to the continuously growing OSM community. It provides an extraordinary geospatial data source for developing web mapping services (e.g., [www.fromtomap.org](www.fromtomap.org)) and conducting related research. The data are collected from the bottom up, owned by no one and updated continuously. For the first time in human history, researchers can acquire an entire world street data set for data-intensive geospatial computing.

Apart from the findings described above, we produced an extraordinary set of data of the natural cities that could be of use to the GIScience community or the science community in general. The natural cities are formed through the interactions of individual nodes, and it is to a great extent a self-organized process. This is a big contrast to the city boundaries extracted from satellite imagery, which is acquired from the top down, costly and inaccurate. We believe that the classified street nodes (or natural cities) constitute a valuable data source for geographic research. The data (1.2 GB) under resolution 600 are available at [http://fromto.hig.se/~bjg/ijgis/Zipf](http://fromto.hig.se/~bjg/ijgis/Zipf) for downloading. The file format is text and each record has the format "ID, Longitude, Latitude, Class". We believe that the natural cities capture well the city pattern of USA. Figure 6 illustrates the top 8% natural cities (equal to 227,803) in the head of the long tail distribution shown in Figure 3.

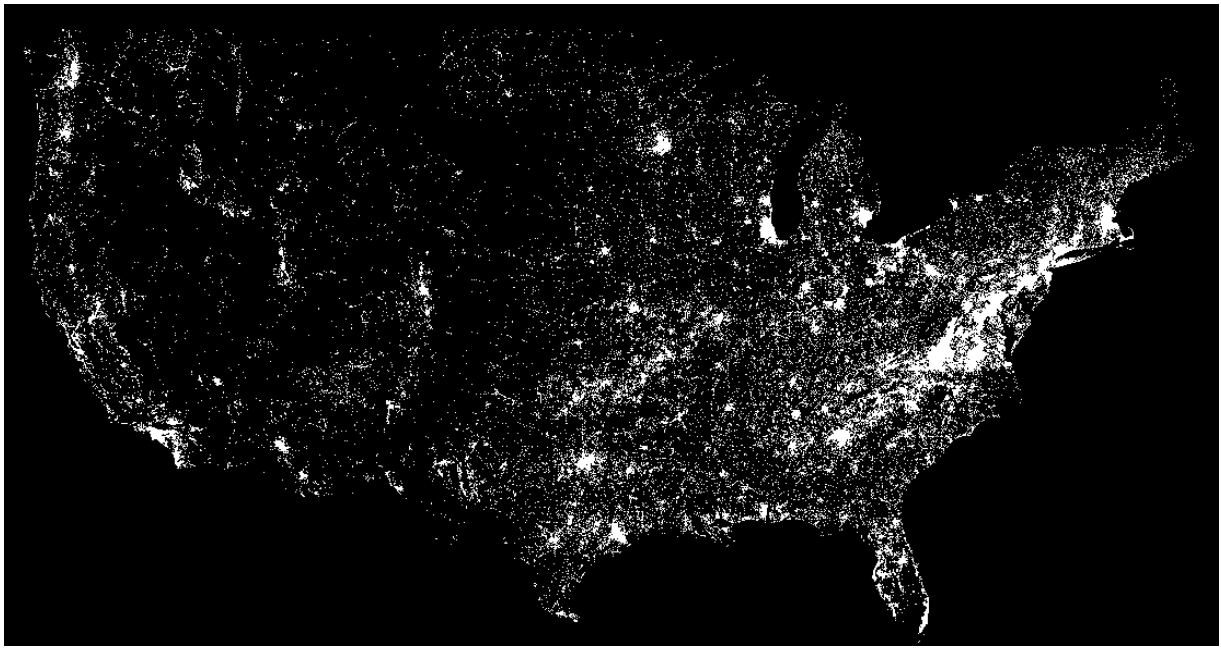

Figure 6: Natural cities in the US (including Alaska and Hawaii but excluded in this visualization) under resolution 600 in the head of the long tail distribution shown in Figure 3

**6. Conclusion**
This paper provides a geospatial perspective to studying the validity of Zipf's law involving all human settlements or population concentrations in the United States. Our investigation leads to the finding that Zipf's law holds remarkably well for the entire country. This implies that cities are power law distributed, and that the Zipf value is around 1.0. However, this does not hold for individual states. Urban areas as a proxy for city sizes are found to be power law distributed, but the Zipf's value is substantially different from 1.0. From this study, we can see a particular advantage of data-intensive computing in scientific discovery. Data and data size do indeed matter! More than half of the entire OSM data is involved in the study.

Apart from the validity of Zipf's law at such a massive data level, this study adds new insights into scaling of geographic space (Jiang 2010). Geographic space is essentially power law distributed. If not, this is because of (1) too small geographic space, in which case we need to expand our scope to include a larger geographic space, or we need to wait a longer time until the geographic space is fully evolved, or (2) a wrong perspective, in which case we need to change our perspective. For example, a street network space based on the connectivity of street segments are never power law distributed, but the one based on the connectivity of streets are (Jiang 2007). This scaling property of geographic space is fundamentally different from the conventional concept of spatial heterogeneity characterized by a Gaussian distribution. Our future work will try to address the issue of implications behind different exponents for different states or countries. We will also seek to tackle the second issue as to why and how this regularity has emerged.

Acknowledgement
We thank the anonymous reviewers for their constructive comments. We also thank Petra Norlund for polishing up our English.

**References:**
Adamic L. A. (2002), Zipf, Power-laws, and Pareto – A ranking tutorial, available at http://www.hpl.hp.com/research/idl/papers/ranking/ranking.html
Auerbach F. (1913), Das Gesetz der Bevölkerungskonzentration, *Petermann's Geographische Mitteilungen*, 59, 74–76.
Batty M. (2006), Rank clocks, *Nature*, 444, 592–596.
Benguigui L. and Blumenfeld-Lieberthal E. (2007), Beyond the power law – a new approach to analyze city size distributions, *Computers, Environment and Urban Systems*, 31, 648 – 666.
Benguigui L., Blumenfeld-Lieberthal, and Czamanski D. (2006), The dynamics of the Tel-Aviv morphology, *Environment and Planning B*, 33.2, 269 – 284.
Borruso G. (2003), Network density and the delimitation of urban areas, *Transactions in GIS*, 7.2, 177 - 191.
Chen Y. (2010), Scaling analysis of the cascade structure of the hierarchy of cities, in: Jiang B. and Yao X. (editors), *Geospatial Analysis and Modeling of Urban Structure and Dynamics*, Springer: Berlin, 91 - 117.
Chen Y. and Zhou Y. (2008), Scaling laws and indications of self-organized criticality in urban systems, *Chaos, Solitons & Fractals*, 35.1, 85 – 98.
Clauset A., Shalizi C. R., and Newman M. E. J. (2009), Power-law distributions in empirical data, *SIAM Review*, 51, 661-703.
Cordoba J.-C. (2008), On the distribution of city sizes, *Journal of Urban Economics*, 63, 177–197.
Eeckhout J. (2004), Gibrat's law for (all) cities, *American Economic Review*, 94, 1429 – 1451.
Gabaix X. and Ioannides Y. M. (2004), The evolution of the city size distributions, in: Henderson V. and Thisse J. F. (eds.), *Handbooks of Regional and Urban Economics*, Elsevier: Oxford, Vol. 4, 2341 – 2378.
Goldstein M. L., Morris S. A., and Yen G. G. (2004), Problems with fitting to the power law distribution, *European Physical Journal B*, 41, 255 – 258.
Goodchild M. (2007), Citizens as sensors: The world of volunteered geography, *GeoJournal*, 69.4, 211-221.
Haklay M. and Weber P. (2008), OpenStreetMap: user-generated street maps, *IEEE Pervasive Computing*, 7(4), 12-18.
Holmes T. J. and Lee S. (2009), Cities as six-by-six-mile squares: Zipf's law? In: Glaeser E. L. (ed.), *The Economics of Agglomerations*, University of Chicago Press: Chicago.
Jacquez, G. M. (2008), Spatial cluster analysis, In: S. Fotheringham and J. Wilson (eds.), *The Handbook of Geographic Information Science*, Blackwell Publishing, 395-416.
Jiang B. (2010), Scaling of geographic space and its implications, A position paper to be presented at *Las Navas 20th Anniversary Meeting on Cognitive and Linguistic Aspects of Geographic Space*, Las Navas del Marques, Avila, Spain, July 5 - 9, 2010.
Jiang B. (2007), A topological pattern of urban street networks: universality and peculiarity, *Physica A: Statistical Mechanics and its Applications*, 384, 647 - 655.
Krugman P. (1996), Confronting the mystery of urban hierarchy, *Journal of the Japanese and International Economies*, 10, 399–418.
Laherrere J. and Sornette D. (1998), Stretched exponential distributions in nature and economy: "fat tails" with characteristic scales, *The European Physical Journal B*, 2.4, 525-539.
Newman M. E. J. (2005), Power laws, Pareto distributions and Zipf's law, *Contemporary Physics*,

46.5, 323-351.

Rozenfeld H. D., Rybski D., Gabaix X. and Makse H. A. (2009), The area and population of cities: new insights from a different perspective on cities, Preprint, http://arxiv.org/abs/1001.5289

Simon H. (1955), On a class of skew distribution functions, *Biometrika*, 42, 425 – 440.

Viswanathan G. M., Buldyrev S. V., Havlin S., da Luz M. G. E., Raposo E. P., and Stanley H. E. (1999), Optimizing the success of random searchers, *Nature*, 401, 911 – 914.

Zipf G. K. (1949), *Human Behavior and the Principles of Least Effort*, Addison Wesley: Cambridge, MA.